\documentclass[a4paper,11pt]{article}

\usepackage{jheppub}

\usepackage{graphicx}

\title{A reanalysis of event shape distributions in electron-positron annihilation}

\author{Zhu-Yu Ren$^{a}$,}
\author{Sheng-Quan Wang$^{a,1}$,\note{Corresponding author.}} \emailAdd{sqwang@alu.cqu.edu.cn}
\author{Jian-Ming Shen$^{b}$,} \emailAdd{shenjm@hnu.edu.cn}
\author{Xing-Gang Wu$^{c}$,} \emailAdd{wuxg@cqu.edu.cn}
\author{Leonardo Di Giustino$^{d,e}$} \emailAdd{leonardo.digiustino@uninsubria.it}

\affiliation{ $^{a}$Department of Physics, Guizhou Minzu University, Guiyang 550025, P.R. China}
\affiliation{ $^{b}$School of Physics and Electronics, Hunan University, Changsha 410082, P.R. China}
\affiliation{ $^{c}$Department of Physics, Chongqing University, Chongqing 401331, P.R. China}
\affiliation{ $^{d}$Department of Science and High Technology, University of Insubria, via valleggio 11, I-22100, Como, Italy}
\affiliation{ $^{e}$INFN, Sezione di Milano-Bicocca, 20126 Milano, Italy}

\abstract{Theoretical calculations for event shape observables are often determined by using the conventional scale setting; i.e. the procedure defined by setting the renormalization scale to the center-of-mass energy $\mu_r=\sqrt{s}$ and evaluating theoretical uncertainties by varying the same scale $\mu_r$ in an arbitrary range. Both the event shape distributions and the extracted QCD coupling $\alpha_s$ are plagued by the large renormalization scale uncertainties when using the conventional scale setting. The Principle of Maximum Conformality (PMC) provides a rigorous method to eliminate the renormalization scheme and scale ambiguities in perturbative QCD predictions. In this paper, we perform a detailed analysis of the event shape observables by applying the PMC method together with the use of the physical $V$-scheme. The PMC scales are not simple single-valued functions, but depend with continuity on the value of the unintegrated event shape variable. This reflects the virtuality of the underlying quark and gluon subprocess and yields to a physical behavior of the scale all over the entire range of each observable. Moreover, the PMC scales in the $V$-scheme exhibits a faster increase compared to the $\overline{\rm MS}$ scheme, and a better convergence in the perturbative series can be obtained. Results obtained by the PMC method for the event shape variables, thrust ($T$), heavy jet mass ($\rho=M^2_H/s$), wide jet broadening ($B_W$), total jet broadening ($B_T$), C-parameter ($C$), are in agreement with the high precision experimental data, and for the case of the jet transition variable $Y_3$, we obtain a first improvement in the results to some extent compared with the $\overline{\rm MS}$ scheme.}


\begin{document}

\maketitle

\flushbottom

\section{Introduction}
\label{sec:1}

Event shape observables in electron-positron annihilation provide an excellent platform for verifying QCD~\cite{Kluth:2006bw}. These observables have expansions of order ${\mathcal O}(\alpha_s)$ at lowest order (LO) of accuracy and thus they are particularly suitable for extracting the strong coupling constant from experimental data~\cite{Dissertori:2009qa,Dissertori:2009ik,Dissertori:2015tfa,Marzani:2019evv}. A primary challenge in achieving accurate predictions with pQCD is to address the uncertainties stemming from theoretical calculations, in particular those related to the renormalization scheme and scale ambiguities. A detailed analysis of these observables can lead to an accurate description of the renormalization scale dependence of the perturbative QCD calculations.

Event shape variables have been extensively studied at the Large Electron Positron (LEP) collider performing high precision measurements~\cite{Abdallah:2003xz, Heister:2003aj, Achard:2004sv, Abbiendi:2004qz,SLD:1994idb}. In order to describe experimental data with high accuracy, it is crucial to achieve the highest possible theoretical precision. Fixed order calculations for event shape observables have been performed up to the NNLO level of accuracy and are shown in Refs.~\cite{GehrmannDeRidder:2007hr, Gehrmann-DeRidder:2008qsl, Weinzierl:2008iv, Weinzierl:2009ms, DelDuca:2016ily,DelDuca:2016csb,Gehrmann:2017xfb}. Results including resummation of infra-red large logarithms have also been performed in Refs.~\cite{Banfi:2004nk,Banfi:2004yd,Banfi:2014sua,Banfi:2016zlc,Chiu:2011qc,Chiu:2012ir,Chien:2010kc,Becher:2012qc,Abbate:2010xh,Hoang:2014wka,Abbate:2012jh,Hoang:2015hka,Benitez:2024nav,Benitez:2025vsp}.

Although a lot of effort has been made to enhance the precision of the theoretical predictions, the choice of the renormalization scale still impacts significantly the results. These uncertainties are fundamental issues of perturbative QCD and are described in detail in Refs.~\cite{Celmaster:1979km,Brodsky:1982gc, DiGiustino:2022ggl}.

According to convention, in the case of event shape observables, one simply sets the renormalization scale to the center-of-mass energy $\sqrt{s}$, this eliminates the logarithmic term $\ln(\mu_r/\sqrt{s})$, and evaluates uncertainties by varying the scale $\mu_r\in[\sqrt{s}/2, 2\sqrt{s}]$. This simplified way of setting the scale has not shown to lead to reliable results in many applications and it is significantly plagued with the well-known renormalization scheme-and-scale ambiguities. For the particular case of the event shape variables fixed order calculations, this method leads to results that underestimate data and that are affected by sizeable uncertainties.

The conventional approach does not take into account the behavior of the QCD perturbative expansion and its convergence that is of asymptotic nature and also affected by the presence of divergent "renormalon" terms at higher orders.  Consequently, the improvement of the precision with the level of accuracy is questionable and strictly process and scale dependent. This affects significantly the theoretical predictions in accuracy and reliability. In order to overcome this obstruction it is necessary to develop a scale-setting method able to eliminate the renormalization scale and scheme ambiguities, leading to a significant improvement of the precision of pQCD theory predictions.

The Principle of Maximum Conformality (PMC)~\cite{Brodsky:2011ta,Brodsky:2012rj,Brodsky:2011ig,Mojaza:2012mf,Brodsky:2013vpa} has been proposed as a procedure that eliminates the dependence on the renormalization scale and scheme. This method stems directly from the Brodsky-Lepage-Mackenzie approach and extends the same method~\cite{Brodsky:1982gc} to all orders in perturbation theory. Fundamental feature of this approach is that it reduces to the Gell-Mann-Low method~\cite{GellMann:1954fq} in the Abelian limit. As it happens for the the Gell-Mann-Low method, the PMC procedure reabsorbs all non-conformal $\beta$ terms into the running coupling constant $\alpha_s$. Differently from the Gell-Mann-Low method, in the PMC procedure this is obtained via a suitable definition of the renormalization scale, $ \mu_{\rm PMC}$, and by a recursive use of the renormalization group equation (RGE) that is applied at all order up to the accuracy of the calculation. This approach starts from first principles and leads to an unambiguous definition of the renormalization scale and to scheme independent predictions. The PMC scale-setting method satisfies all the requirements of consistency given by renormalization group invariance~\cite{Brodsky:2012ms,Wu:2014iba,Wu:2019mky,DiGiustino:2023jiq}. Moreover, the elimination of the $\beta$ terms from the perturbative calculation cancels the renormalon growth and it improves the convergence of the perturbative QCD series. Thus, the PMC scale is independent of the initial choice of the scale and the precision of the theoretical predictions is greatly improved by the elimination of the scale-and-scheme uncertainties.

At present, the PMC method has been successfully applied to eliminate the uncertainty related to the renormalization scale for the event shape variables in the conventional $\overline{\rm MS}$-scheme~\cite{Wang:2019ljl,Wang:2019isi,Wang:2021tak}. The resulting PMC scales are not a fixed value, but change with event shape variable. The PMC predictions agree with precise event shape distributions measured at the LEP experiment and reliable coupling constants in the conventional $\overline{\rm MS}$-scheme are extracted. However, in the case of conventional scale setting, the NLO and NNLO QCD contributions for event shape variables are large and thus the pQCD series show a slow convergence in the $\overline{\rm MS}$-scheme. In addition, the determined scales are relatively small in the $\overline{\rm MS}$-scheme, which lead to a large QCD contribution, especially for the jet-transition variable $Y_3$, the prediction deviates substantially from the experimental data due to the small scale~\cite{Gehrmann:2014uva}.

A proper choice of physical scheme is helpful to improve the pQCD convergence and to avoid small scale problem. The $V$-scheme given by the static potential between two heavy quarks~\cite{Appelquist:1977tw,Fischler:1977yf,Schroder:1998vy}, which is particularly suitable to obtain a more physical definition of the strong coupling~\cite{Brodsky:1982gc} defined by the effective-charge. It is noted that the PMC scale in QED is identical to the QCD PMC scale in the physical $V$-scheme~\cite{Wang:2020ckr}. Fundamental feature of the $V$-scheme is the gauge invariance at all orders. Results for several application of the $V$-scheme are available in the literature, where this scheme has been successfully applied in several phenomenological QCD studies~\cite{Brodsky:1995ds,Brodsky:1997dh,Brodsky:1998mf,Yu:2021yvw,Zhou:2022fyb}. As a step forward, in this paper, we give a detailed analysis of the event shape observables by applying the PMC method together with the use of the $V$-scheme. The PMC predictions are scheme independent, which are explicitly ensured by the ``commensurate scale relations" (CSR)~\cite{Brodsky:1994eh, Lu:1992nt}.

The remaining sections of this paper are organized as follows. We present in Sec.~\ref{sec:2} the calculation technology for applying the PMC method to event shape observables in electron-positron annihilation; in Sec.~\ref{sec:3}, we give theoretical predictions and discussions; Sec.~\ref{sec:4} is reserved for a summary.

\section{PMC scale-setting for event shape distributions}
\label{sec:2}
Using the conventional $\overline{\rm MS}$-scheme, the perturbative expansion for the distribution of an event shape observable $y$ up to NLO at the center-of-mass energy $\sqrt{s}$ is given by:
\begin{eqnarray}
\frac{1}{\sigma_0}\frac{d\sigma}{dy}&=&A(y)\,a_s(\sqrt{s})+B(y)\,a^2_s(\sqrt{s}))+{\cal O}(a^3_s), \label{eq:one}
\end{eqnarray}
where $a_s(\sqrt{s})=\alpha_s(\sqrt{s})/(2\pi)$ and $\sigma_0$ is the born cross-section for $e^+e^- \to q\bar q$. The $A(y)$ is the LO coefficient and the $B(y)$ is the NLO coefficient. Event shape distributions are usually normalized to the total hadronic cross section:
\begin{eqnarray}
\sigma_{tot}=\sigma_0\left(1+\frac{3}{2}\,C_F\,a_s(\sqrt{s})+{\cal O}(a^2_s)\right),
\end{eqnarray}
where the color factor $C_F$=4/3.
Thus, the normalized event shape observable in Eq.(\ref{eq:one}) becomes:
\begin{eqnarray}
\frac{1}{\sigma_{tot}}\frac{d\sigma}{dy}&=&\bar{A}(y)\,a_s(\sqrt{s})+\bar{B}(y)\,a^2_s(\sqrt{s})+{\cal O}(a^3_s).
\label{eventshapet}
\end{eqnarray}
The perturbative coefficients $\bar{A}(y)$ and $\bar{B}(y)$ are given by $\bar{A}(y)=A(y)$, $\bar{B}(y)=B(y)-3/2\,C_F\,A(y)$. These coefficients are calculated at the scale $\mu_r=\sqrt{s}$; the general renormalization scale $\mu_r$ dependence can be recovered via RGE transformations leading to the perturbative coefficients: $\bar{A}(y,\mu_r)$, $\bar{B}(y,\mu_r)$.

Keeping track of the contributions related to each particular color factor, we can separate the NLO coefficient, $\bar{B}(y,\mu_r)$, into two parts, such as:
\begin{eqnarray}
\bar{B}(y,\mu_r)=\bar{B}(y,\mu_r)_{\rm in}+\bar{B}(y,\mu_r)_{n_f}\cdot n_f, \label{Eq:BMS}
\end{eqnarray}

The relation of the coupling constant between the $\overline{\rm MS}$ scheme and the $V$ scheme is given by:
\begin{eqnarray}
a_s(\mu_r) = \sum_{i=1}^{\infty} r^{V}_{i} a_s^{V, i}(\mu_r),
\label{amscon}
\end{eqnarray}
the first two coefficients are~\cite{Anzai:2009tm,Smirnov:2009fh,Smirnov:2008pn,Kataev:2015yha,Kataev:2023sru}
\begin{eqnarray}
r^{V}_1&=&1, \nonumber\\
r^{V}_2&=&-\frac{31}{18}C_A+\frac{20}{18}T_F\cdot n_f.
\label{eventshapetp}
\end{eqnarray}
We can transform the $\overline{\rm MS}$ scheme to the $V$ scheme by using Eq.(\ref{amscon}), and Eq.(\ref{eventshapet}) becomes:
\begin{eqnarray}
\frac{1}{\sigma_{tot}}\frac{d\sigma}{dy}=\bar{A}(y,\mu_r)^V\,a^V_s(\mu_r)+\bar{B}(y,\mu_r)^V\,a^{V,2}_s(\mu_r)+{\cal O}(a^{V,3}_s),
\label{eventshapets}
\end{eqnarray}
following Eq.(\ref{Eq:BMS}) for the NLO coefficient , $\bar{B}(y,\mu_r)^V$ in the $V$ scheme can be written in the same form:
\begin{eqnarray}
\bar{B}(y,\mu_r)^V=\bar{B}(y,\mu_r)^V_{\rm in}+\bar{B}(y,\mu_r)^V_{n_f}\cdot n_f.
\end{eqnarray}
According to the PMC scale-setting method, all non-conformal $\beta$ terms are reabsorbed into the running coupling constant $\alpha_s^V$, by a shift of the scale, achieving the following final conformal form:
\begin{eqnarray}
\frac{1}{\sigma_{tot}}\frac{d\sigma}{dy}=\bar{A}(y,\mu_r)^V\,a^V_s(Q_*)+\bar{B}(y,\mu_r)^V_{\rm con}\,a^ {V,2}_s(Q_*)+{\cal O}(a^{V,3}_s),
\label{eventshapety}
\end{eqnarray}
where $Q_*$ stands for the PMC scale, given at LO by:
\begin{eqnarray}
\ln\frac{Q_*}{\mu_r}=\frac{\,3\bar{B}(y,\mu_r)^V_{n_f}}{2\bar{A}(y,\mu_r)^V}+{\cal O}(a_s),
\label{eventshapett}
\end{eqnarray}
while the conformal coefficient is given by:
\begin{eqnarray}
\bar{B}(y,\mu_r)^V_{\rm con}=\frac{33}{2}\bar{B}(y,\mu_r)^V_{n_f}+\bar{B}(y,\mu_r)^V_{\rm in}.
\label{evenPMCcon}
\end{eqnarray}

Since all non-conformal terms cancel and only conformal terms remain in Eq.(\ref{eventshapety}). Conformal coefficients are scale-invariant, so are the PMC scales, we then obtain a scale invariant form that is no longer affected by
scale ambiguities and the pQCD predictions are free from renormalization scale uncertainties.

\section{Theoretical predictions and discussions}
\label{sec:3}

Numerical calculations for the NLO perturbative coefficient, have been performed using the EVENT2 program~\cite{Catani:1996jh}. The NNLO perturbative coefficients have been obtained by using the EERAD3 program~\cite{Ridder:2014wza}, and they have been checked using the results of Ref.\cite{Weinzierl:2009ms}. We adopt two-loop QCD coupling constant, and its asymptotic scale $\Lambda^{\overline{MS}}_{\rm QCD}$ is determined by the world average value $\alpha_s(M_Z)=0.1180$~\cite{ParticleDataGroup:2024cfk}. The asymptotic scale in the $V$ scheme is obtained by $\Lambda^{V}_{\rm QCD}=\Lambda^{\overline{MS}}_{\rm QCD}\,\exp[(31/6-(5/9)\,n_f)/\beta_0]$ where $\beta_0=11-2/3\,n_f$.

In this paper, we show results for the differential distributions of the following event shape variables: thrust ($T$), heavy jet mass ($\rho=M^2_H/s$), wide jet broadening ($B_W$), total jet broadening ($B_T$), C-parameter ($C$) and jet-transition  ($Y_3$). These event shape distributions have been measured at LEP by several experiments and at different center-of-mass energies ranging from $35$ to $206$ GeV~\cite{Heister:2003aj}.

\subsection{Event shape distributions at the $Z^0$ peak using conventional scale setting }

\begin{figure} [htb]
\centering
\includegraphics[width=0.45\textwidth]{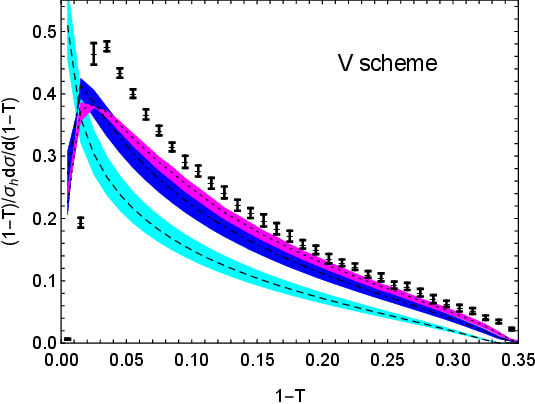}
\includegraphics[width=0.45\textwidth]{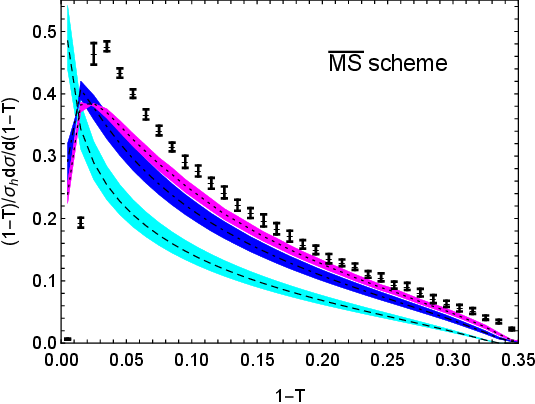}
\caption{The thrust ($1-T$) differential distributions using the conventional scale-setting method at $\sqrt{s}=91.2$ GeV in the $V$ scheme and the $\overline{\rm MS}$ scheme. The dashed, dotdashed and dotted lines are the conventional results at LO, NLO and NNLO, respectively. The bands for the theoretical predictions are obtained by varying $\mu_r\in[\sqrt{s}/2,2\sqrt{s}]$. The experimental data are taken from the ALEPH Collaboration~\cite{Heister:2003aj}.}
\label{TdistCon}
\end{figure}

Theoretical predictions for the event shape distributions using the conventional scale-setting, show similar results if one adopts either the $\overline{\rm MS}$ scheme or the $V$ scheme. We show in Fig.(\ref{TdistCon}) the case of thrust distributions in the two schemes as an example, using the conventional approach with the scale set at the center-of-mass energy $\sqrt{s}=91.2$ GeV. The error bands for the theoretical predictions are obtained by varying $\mu_r\in[\sqrt{s}/2,2\sqrt{s}]$. It is noted that the convergence of the perturbative series improves to a certain degree by using the $V$ scheme with respect to the $\overline{\rm MS}$ scheme. However, pQCD predictions for both the $V$ scheme and the $\overline{\rm MS}$ scheme, still significantly underestimate the precise experimental data. The NLO results do not overlap with LO predictions; for the $\overline{\rm MS}$ scheme, the NNLO results still do not overlap with NLO predictions, while for the $V$ scheme, the NNLO results have only a partial overlap with NLO predictions in the intermediate region. Thus, the method of estimating unknown higher order terms by simply setting the scale to the center-of-mass energy and varying the scale in the range $\mu_r\in[\sqrt{s}/2, 2\sqrt{s}]$, does not lead to reliable results. In fact, the errors on predictions, depend strictly on the particular value of the chosen scale and to the range of its variation. The latter is related to the non-conformal $\beta$ terms~\cite{Beneke:1994qe,Neubert:1994vb,Beneke:1998ui} and not to the conformal terms. Besides, we do not know {\it a priori} what is the correct range to obtain a reasonable prediction for the theoretical error~\cite{Wu:2013ei}.

\subsection{Event shape distributions at the $Z^0$ peak using PMC scale setting }

\begin{figure} [htb]
\centering
\includegraphics[width=0.45\textwidth]{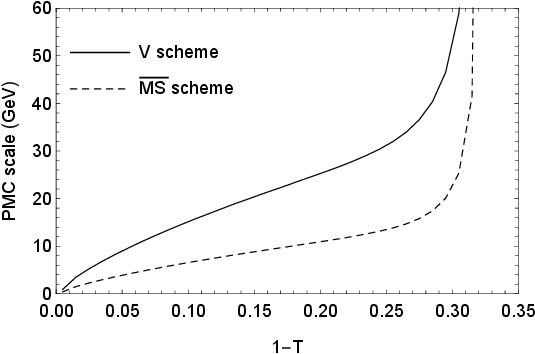} 
\includegraphics[width=0.45\textwidth]{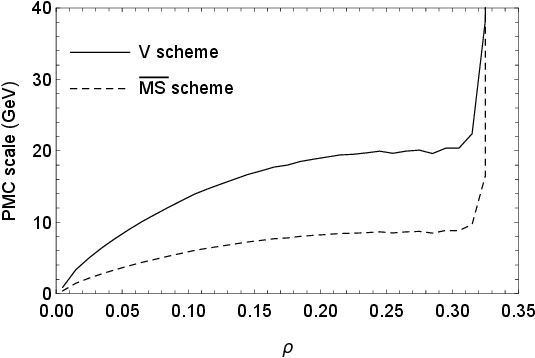} 
\includegraphics[width=0.45\textwidth]{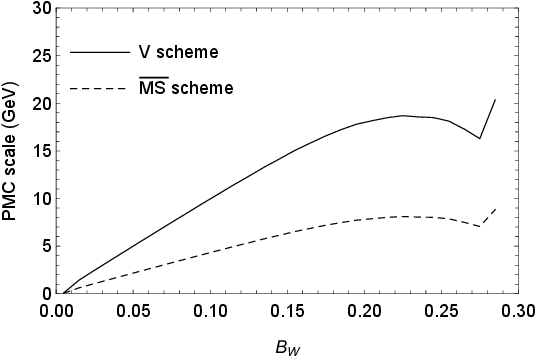} 
\includegraphics[width=0.45\textwidth]{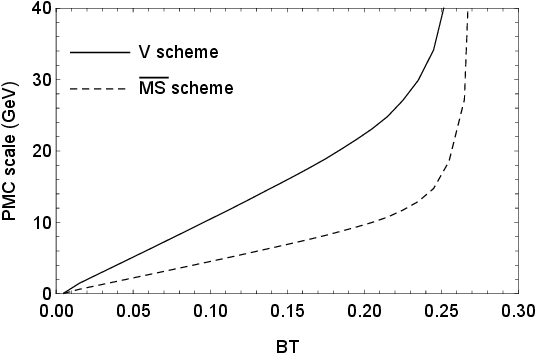} 
\includegraphics[width=0.45\textwidth]{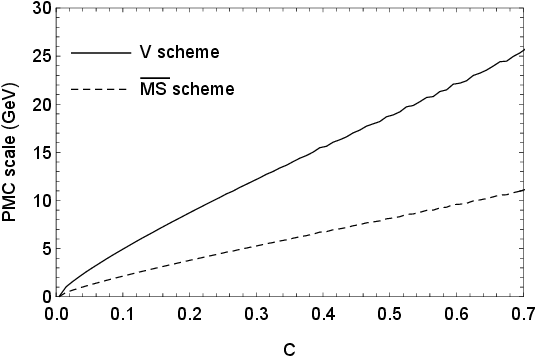} 
\caption{PMC scales at $\sqrt{s}=91.2$ GeV in the $V$ scheme (solid line) and in $\overline{\rm MS}$ scheme (dashed line) for the event shape variables: thrust ($1-T$), heavy jet mass ($\rho$), wide jet broadening ($B_W$), total jet broadening ($B_T$) and C-parameter ($C$) respectively.}
\label{eventshapepmcscale}
\end{figure}

By introducing the PMC method, the renormalization scales are determined by reabsorbing the non-conformal $\beta$ terms into the coupling constant. In Fig.(\ref{eventshapepmcscale}) we display the PMC scales at $\sqrt{s}=91.2$ GeV for the five event shape observables, that consist of thrust ($T$), heavy jet mass ($\rho=M^2_H/s$), wide jet broadening ($B_W$), total jet broadening ($B_T$) and the C-parameter ($C$) in the $V$ scheme. We also show the PMC scales in the $\overline{\rm MS}$ scheme as a comparison.

As expected, PMC scales in the $V$ scheme have a faster increasing behavior with the increasing of the event shape variable, if compared to the $\overline{\rm MS}$ scheme. This reflects the size of the terms related to each color factor that readjust at each order by transforming to the particular definition of the $V$ scheme, that consequently leads to larger scales with respect to the $\overline{\rm MS}$ scheme. Additionally, the scales in both schemes are significantly smaller than the center-of-mass energy of the process, it follows a greater coupling value and an increased distribution. Unlike the conventional scale-setting method, the PMC scale is not a single-valued function but it changes dynamically with the value of the event shape observable. This behavior reflects the virtuality of the quark and gluon subprocess, making the PMC scale consistent with the correct physical soft limit of the scale at the left-boundary. The PMC scale correctly interpolates all dynamical regions of the event shape variables determining where non-perturbative effects of  QCD dynamics become more important. On the contrary, the conventional scale, set to the center-of-mass energy $\sqrt{s}$,
is lacking of a thorough description and control of the dynamics with respect to the physical range of the selected observable and the occurrence of the spoiling non-perturbative terms is experienced unpredictably during calculations.
In fact, in two jet region (i.e. the left-boundary), the PMC scale approaches a soft limit, revealing the necessity of including non-perturbative effects, such as  resummation of the IR-large logarithms related to soft-and-collinear divergences in order to improve the perturbative results and the description of the physics. This is consistent with soft-collinear effective theories (e.g. SCET) that predict soft scales in the two jet regions~\cite{Bauer:2000ew,Bauer:2000yr,Bauer:2001yt,Beneke:2002ph,DiGiustino:2011jn,Aglietti:2025jdj}.

The soft-limit of the PMC in the two-jet region is not due to an ad-hoc introduction of a particular kinematic cutoff in the PMC calculation, but it is a direct consequence of the $N_f$-terms that stem from the loop-integration and thus it is directly connected to the loop kinematic boundary. In fact the PMC-scale dynamics reflects the dynamics of the virtuality of the gluons and quarks in the loops, this virtuality is constrained by kinematics and thus it leads to the zero-limit in the two-jet region. We remind that PMC methodology applies to the perturbative QCD calculation and thus the validity range of the PMC-scale is determined by the $Q_* \gg \Lambda_{\rm QCD}$ condition that is verified in the following bin-range of each variable at $\sqrt{s}=91.2$ GeV: $0.01 \leq (1-T) \leq 0.33$, $0.02 \leq \rho \leq 0.33$,  $0.04 \leq B_W \leq 0.29$, $0.04 \leq B_T \leq 0.29$, $0.06  \leq  C  \leq 0.75$ where the PMC-scale is: $Q_* > 1.5 $ GeV and thus in the perturbative region.

\begin{figure} [htb]
\centering
\includegraphics[width=0.45\textwidth]{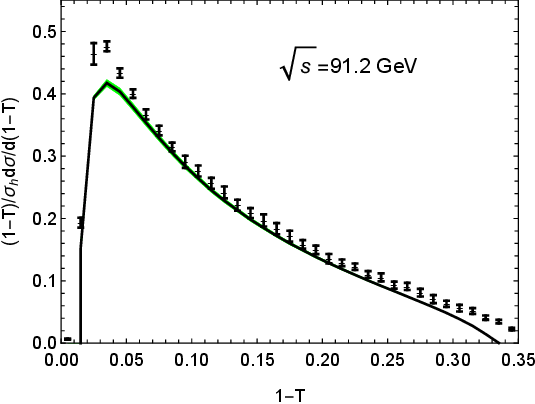} 
\includegraphics[width=0.45\textwidth]{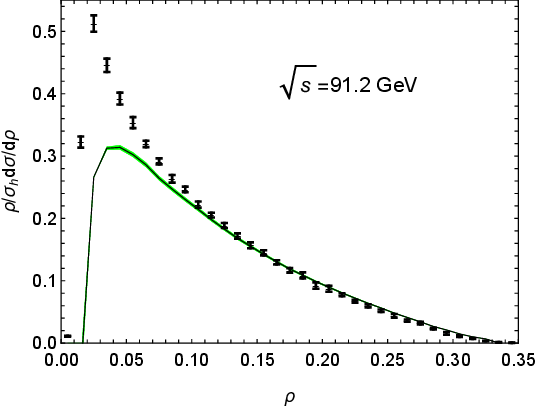} 
\includegraphics[width=0.45\textwidth]{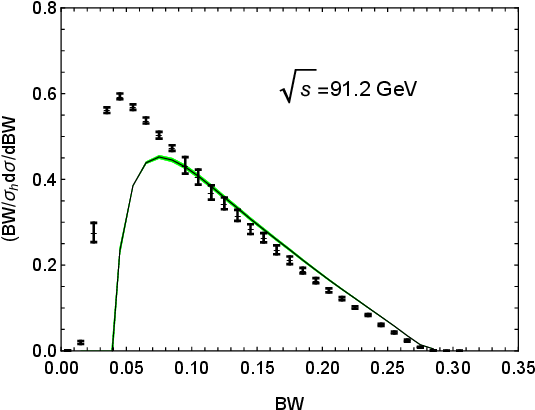} 
\includegraphics[width=0.45\textwidth]{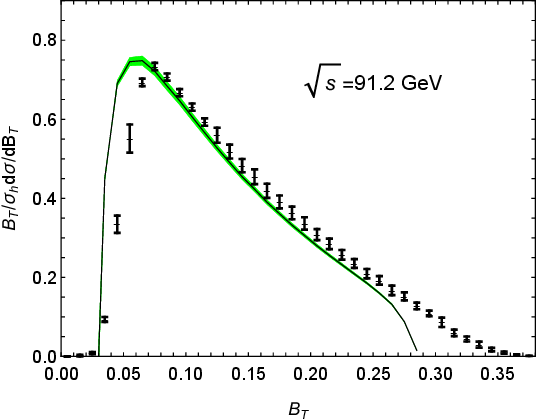} 
\includegraphics[width=0.45\textwidth]{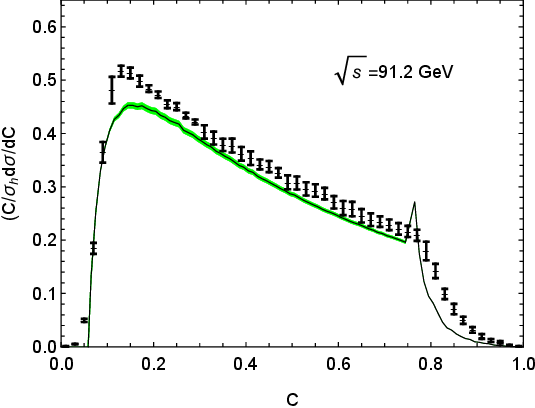} 
\caption{Results for the event shape variables:  thrust ($1-T$), heavy jet mass ($\rho$), wide jet broadening ($B_W$), total jet broadening ($B_T$) and C-parameter ($C$) distributions, using the PMC scale-setting at $\sqrt{s}=91.2$ GeV. Error bands related to the uncertainty $\Delta\alpha_s(M_Z)=\pm0.0009$~\cite{ParticleDataGroup:2024cfk} are also shown. The experimental data are taken from the ALEPH Collaboration~\cite{Heister:2003aj}. }
\label{eventshapePMC}
\end{figure}

Moreover, the behavior of PMC conformal coefficients is different from the coefficients determined by the conventional scale-setting method at each order of accuracy and results obtained by PMC increase all over the kinematic regions of the event shape variables, with respect to the conventional scale-setting method. More explicitly, in Fig.(\ref{eventshapePMC}) we give the results of event shape distribution at $\sqrt{s}=91.2$ GeV for thrust ($T$), heavy jet mass ($\rho=M^2_H/s$), wide jet broadening ($B_W$), total jet broadening ($B_T$) and the C-parameter ($C$) in the $V$ scheme. The PMC method eliminates scale uncertainties, since both PMC scales and conformal coefficients are independent of renormalization scale choice; the uncertainty for the PMC predictions are obtained by $\Delta\alpha_s(M_Z)=\pm0.0009$~\cite{ParticleDataGroup:2024cfk}. The event shape distributions have been measured experimentally with a high degree of accuracy at $\sqrt{s}=91.2$ GeV. Results for the same event shape variables at $\sqrt{s}=91.2$ GeV in the $\overline{\rm MS}$ scheme have been provided in literature by previous analysis~\cite{Gehrmann:2014uva, Wang:2021tak}. As shown in Fig.(\ref{eventshapePMC}), the PMC predictions are in good agreement with the experimental data.
And in fact, the $V$ scheme leads to a larger scale with respect to the $\overline{\rm MS}$ scheme. This scale plays a crucial role in determining the accuracy of the predictions. A larger scale can lead to a more rapid increase in values, better capturing the characteristics of the physical process. In contrast, a slowly rising energy scale may fail to adequately represent the nuances of the physical process, potentially resulting in less accurate predictions. Therefore, it is essential to carefully consider the scale when conducting theoretical calculations to ensure the accuracy and reliability of the results.

Besides, we also notice clear deviations from data of the five event shapes observables in both the two-jet and the multijet regions. These are due to the  presence of large IR-logarithms that are enhanced in these regions spoiling the perturbative fixed-order QCD predictions. It is possible to restore the correct perturbative regime by resumming the logarithmically enhanced terms to all orders in the coupling constant. The resummation of large logarithms has been extensively studied in the literature. In fact, due to kinematical constraints, the event shape distribution at LO is limited to a certain range. For instance, the thrust distribution is restricted to the range of $0\leq(1-T)\leq1/3$ and the $C$-parameter distribution is restricted to the range of $0\leq C\leq0.75$. Hence, in this region, PMC predictions are dominated by the 3-jet process contributions and thus show a better agreement with the accurate experimental data. Outside this range, pQCD calculations are significantly affected by missing higher order contributions and thus the 3-jet process contribution alone cannot thoroughly describe data. As reported in literature, in order to obtain a correct analysis of event shape variables performed using the conventional method, one has to include also the resummation of large logarithms outside the kinematical constrained range. Furthermore, in addition to the resummation of large logarithms, non-perturbation effects should be considered in the two-jet region.

We remark that a partial resummation of the IR large-logarithms already occur in the perturbative calculations up to NNLO. In fact, this explains the presence of the peaks in the distributions and  in the two-jet region. This partial resummation is not related to the PMC, but is simply due to the improvement of the precision with the order of accuracy and the occurrence of the IR-large-logarithms at higher order resum with those of the lower orders. This partial resummation is incomplete, since the series is truncated at the NNLO. Thus, it cannot restore the correct behavior of the distributions in the region on left-side of the peak close to the left-boundary, where instead the distributions oscillate severely. Thus, in this region the entire resummation procedure is required in order to restore the correct predictive behavior of the perturbative calculation. Besides, in Ref.~\cite{DiGiustino:2021nep}, it was also shown that, even though the presence of the peak is due to resummation effects, its position is more related to the renormalization scale and in particular to the PMC scale. Thus, the use of the PMC can improve the perturbative theoretical predictions, extending the range of validity of the fixed-order calculation and the comparison with data. The PMC scale itself can indicate the dynamical range of validity of the perturbative calculation and the region where also other effects such as the resummation of the IR-large-logarithms must be taken into account, in order to have a complete and correct description of the physics all over the entire physical range\footnote{A more detailed study of the interplay between PMC and Resummation for event shape variables is in progress.}.

\begin{figure} [htb]
\centering
\includegraphics[width=0.50\textwidth]{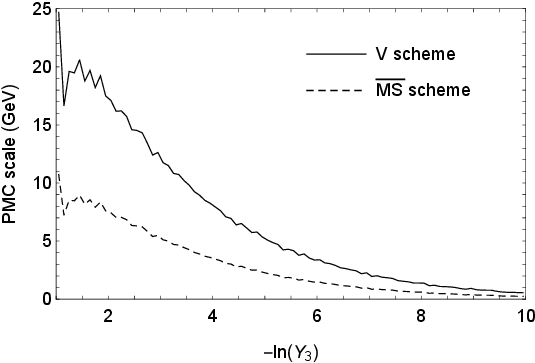}
\caption{Results for the PMC scales  in the $V$ scheme (solid line) and in the $\overline{\rm MS}$ scheme (dashed line) for the jet-transition variable $Y_3$ at the center-of-mass energy $\sqrt{s}=91.2$ GeV. }
\label{y3PMCscale}
\end{figure}

As mentioned in Ref.\cite{Gehrmann:2014uva}, results for the jet-transition variable $Y_3$ deviate substantially from the experimental data. We have also calculated the jet-transition variable $Y_3$ in the $V$ scheme. The PMC scales for the jet transition variable $Y_3$ are presented in Fig.(\ref{y3PMCscale}). This shows that  PMC scales in the $\overline{\rm MS}$ scheme are very small, especially in the region of $5\leq(-lnY_3)\leq10$. The PMC scales in the $V$ scheme have larger values with respect to the PMC scales in the $\overline{\rm MS}$ scheme.

\begin{figure} [htb]
\centering
\includegraphics[width=0.50\textwidth]{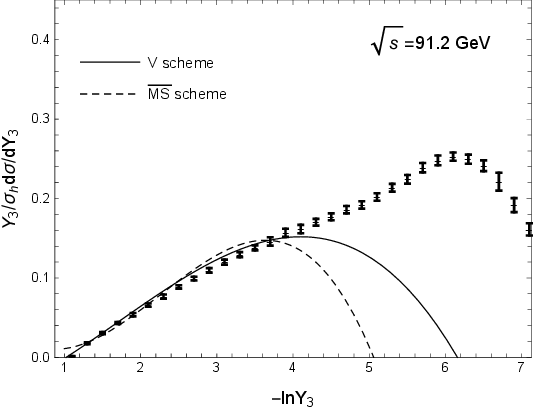}
\caption{Results for the jet-transition $Y_3$-distribution at the center-of-mass energy $\sqrt{s}=91.2$ GeV, using the PMC method in the $V$ scheme (solid line) and in the $\overline{\rm MS}$ scheme (dashed line). Experimental data are taken from the ALEPH Collaboration~\cite{Heister:2003aj}. }
\label{eventshapeY3PMC}
\end{figure}

In Fig.(\ref{eventshapeY3PMC}) we present the jet-transition $Y_3$ distribution using the $V$ scheme and the $\overline{\rm MS}$ scheme for $\sqrt{s}=91.2$ GeV. This shows that the results of the jet transition variable $Y_3$ improve with the $V$ scheme, even though there is still discrepancy between the theoretical predictions and the experimental data in the region of the experimental peak. In this region, distributions obtained by the 3-jet process contributions do not describe exhaustively the physics around the peak and other contributions should be taken into account, such as resummation of large infrared logarithms of the type $\alpha^n_s\ln^m(Y_3)$ .

\subsection{Event shape distributions above and below the $Z^0$ peak }

\begin{figure} [htb]
\centering
\includegraphics[width=0.45\textwidth]{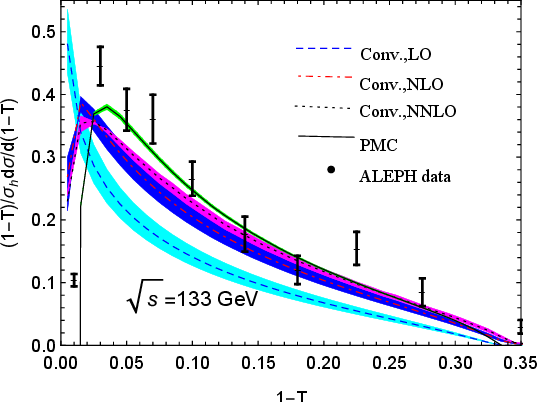} 
\includegraphics[width=0.45\textwidth]{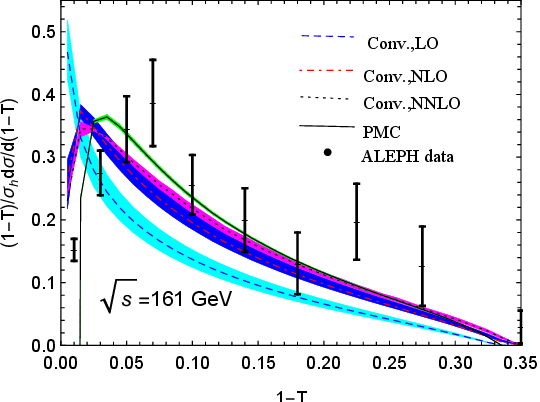} 
\includegraphics[width=0.45\textwidth]{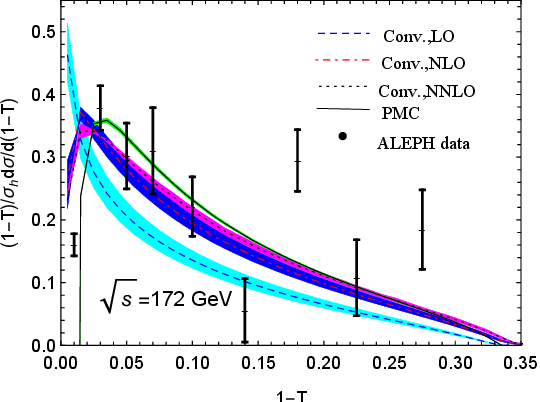} 
\includegraphics[width=0.45\textwidth]{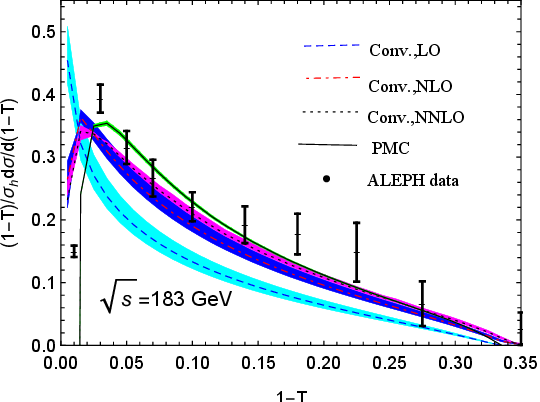} 
\includegraphics[width=0.45\textwidth]{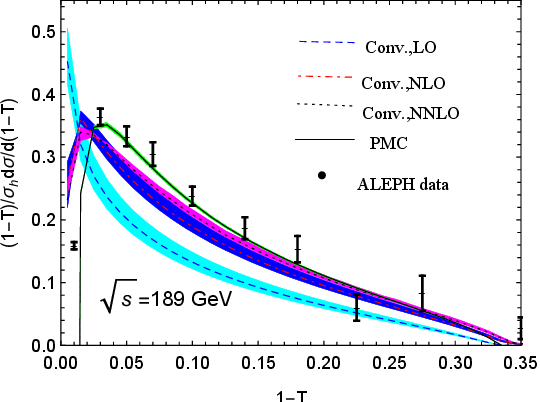} 
\includegraphics[width=0.45\textwidth]{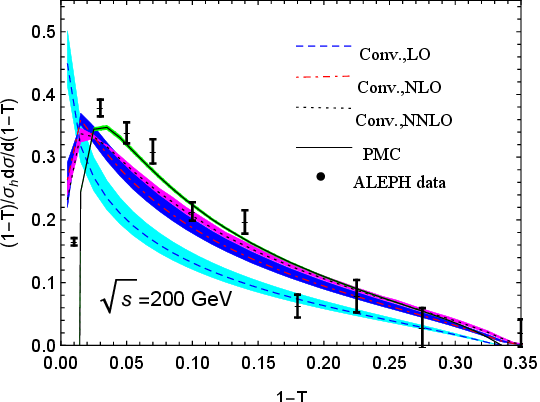}
\includegraphics[width=0.45\textwidth]{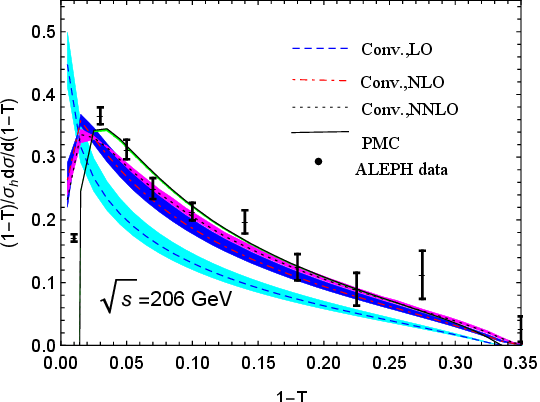}
\caption{Results for thrust distribution using the conventional scale-setting (Conv.) at LO (dashed), NLO (dotdashed ) and NNLO (dotted)~\cite{GehrmannDeRidder:2007hr} and results for the PMC scale-setting (solid line) are shown for $\sqrt{s}=133$, $161$, $172$, $183$, $189$, $200$, $206$ GeV respectively.. Error bands for the conventional predictions are obtained by varying $\mu_r\in[\sqrt{s}/2,2\sqrt{s}]$, while error bands for the PMC predictions are given by the uncertainty $\Delta\alpha_s(M_Z)=\pm0.0009$~\cite{ParticleDataGroup:2024cfk}. The experimental data are taken from ALEPH~\cite{Heister:2003aj}.}
\label{eventshapeTPMC}
\end{figure}

\begin{figure} [htb]
\centering
\includegraphics[width=0.45\textwidth]{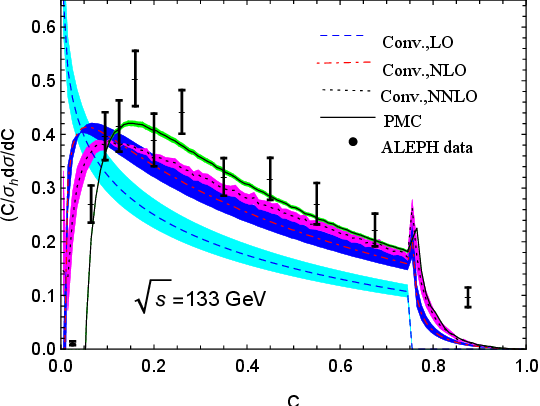} 
\includegraphics[width=0.45\textwidth]{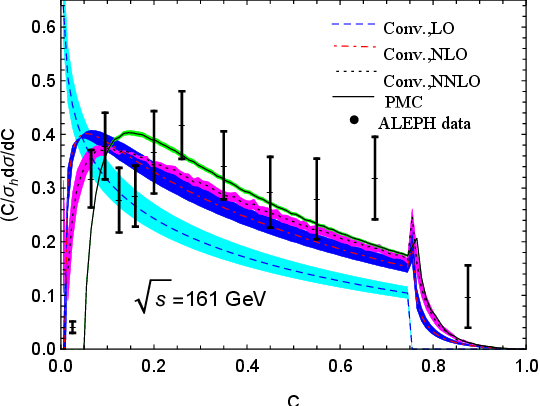} 
\includegraphics[width=0.45\textwidth]{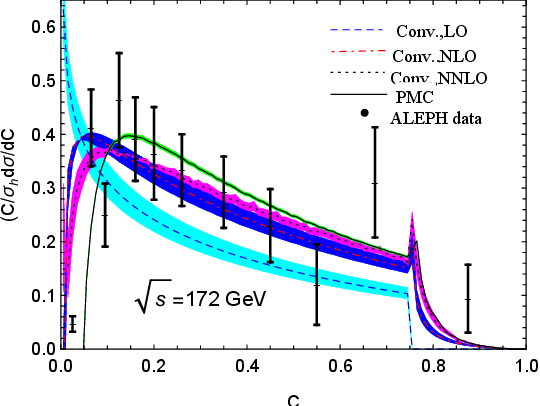} 
\includegraphics[width=0.45\textwidth]{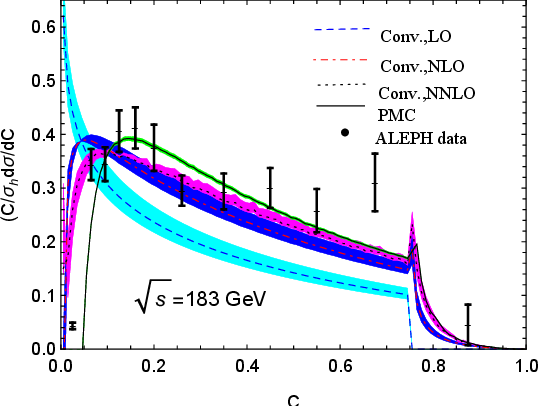} 
\includegraphics[width=0.45\textwidth]{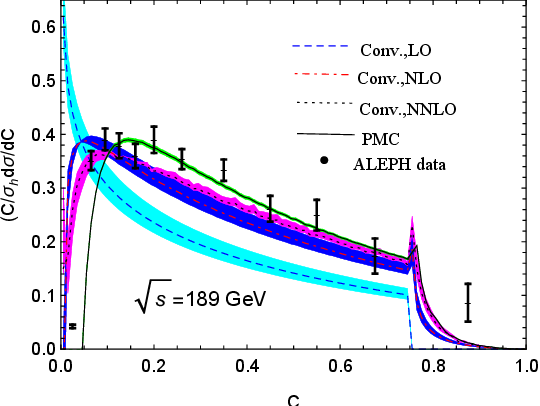} 
\includegraphics[width=0.45\textwidth]{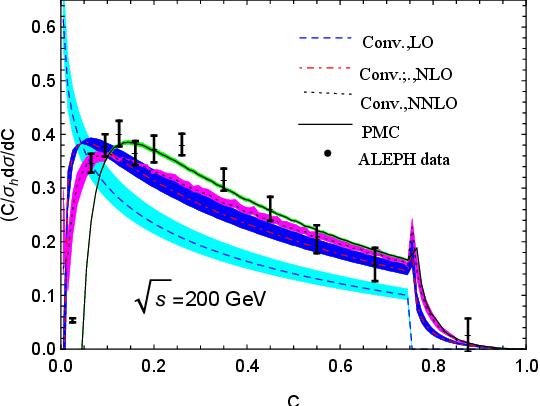}
\includegraphics[width=0.45\textwidth]{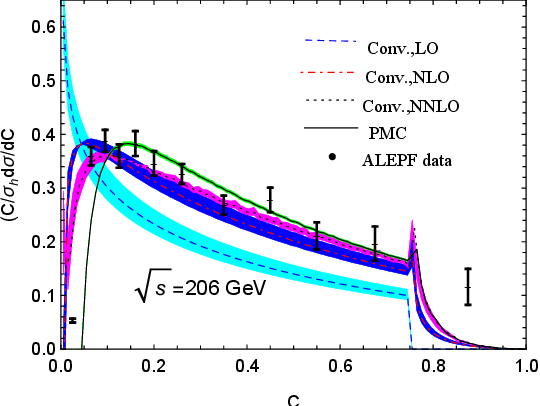}
\caption{Results for C-parameter ($C$) distribution using the conventional scale-setting (Conv.) at LO (dashed), NLO (dotdashed ) and NNLO (dotted)~\cite{GehrmannDeRidder:2007hr} and results for the PMC scale-setting (solid line) are shown for $\sqrt{s}=133$, $161$, $172$, $183$, $189$, $200$, $206$ GeV respectively. Error bands for the conventional predictions are obtained by varying $\mu_r\in[\sqrt{s}/2,2\sqrt{s}]$, while error bands for the PMC predictions are given by the uncertainty $\Delta\alpha_s(M_Z)=\pm0.0009$~\cite{ParticleDataGroup:2024cfk}. The experimental data are taken from ALEPH~\cite{Heister:2003aj}.}
\label{eventshapeCPMC}
\end{figure}

In addition to the event shape distributions at the center-of-mass energy $\sqrt{s}=91.2$ GeV, we also present results for event shape distributions at higher energies. Results for thrust ($T$) and C-parameter ($C$) distributions are shown in Figs.(\ref{eventshapeTPMC}) and (\ref{eventshapeCPMC}) respectively, including the comparison between the conventional and the PMC results. Results for each event shape variable, at different energies, i.e. $\sqrt{s}=133$, $161$, $172$, $183$, $189$, $200$, $206$ GeV,  are shown together with their experimental data measured with high accuracy at the LEP facility~\cite{Abdallah:2003xz}. In Figs.(\ref{eventshapeTPMC}) and (\ref{eventshapeCPMC}), we notice that the PMC results for both the thrust ($T$) and the C-parameter ($C$) have been improved at $\sqrt{s}=133$, $161$, $172$, $183$, $189$, $200$, $206$ GeV. Given the large uncertainties of the experimental data at $\sqrt{s}=161$, $172$ GeV,  both the conventional and the PMC results lead to analogous predictions if compared to the experimental data. While at the energy of $\sqrt{s}=133$, $183$, $189$, $200$, $206$ GeV,  PMC results have a better fit with the experimental data if compared to the conventional results.

\begin{figure} [htb]
\centering
\includegraphics[width=0.45\textwidth]{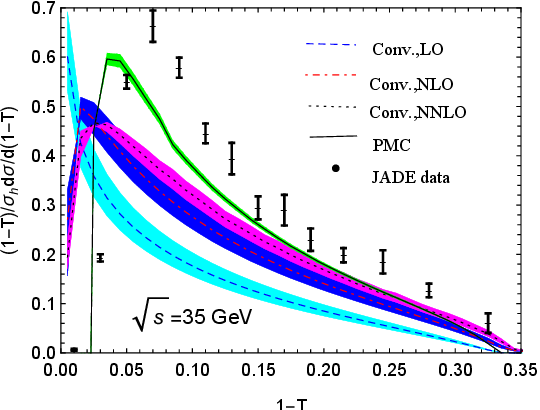}
\includegraphics[width=0.45\textwidth]{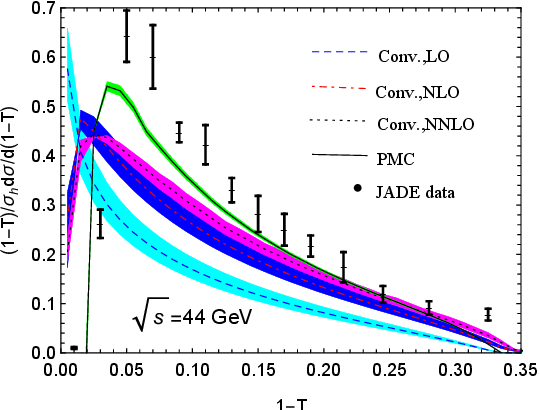}
\caption{ Results for thrust distribution using the conventional scale-setting (Conv.) at LO (dashed), NLO (dotdashed ) and NNLO (dotted)~\cite{GehrmannDeRidder:2007hr} and results for the PMC  scale-setting (solid line) are shown for $\sqrt{s}=35$, $44$ GeV respectively.
Error bands for the conventional predictions are obtained by varying $\mu_r\in[\sqrt{s}/2,2\sqrt{s}]$. Error bands for PMC predictions are given by the uncertainty $\Delta\alpha_s(M_Z)=\pm0.0009$~\cite{ParticleDataGroup:2024cfk}. The experimental data are taken from JADE~\cite{MovillaFernandez:1997fr,Biebel:1999zt}.}.
\label{eventshapeTPMC35}
\end{figure}

\begin{figure} [htb]
\centering
\includegraphics[width=0.45\textwidth]{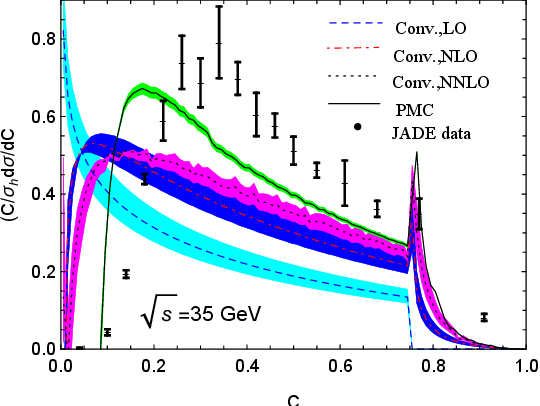}
\includegraphics[width=0.45\textwidth]{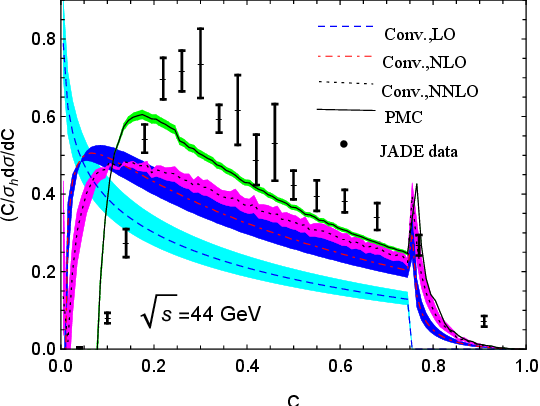}
\caption{Results for the C-parameter ($C$) distribution using the conventional scale-setting (Conv.) at LO (dashed), NLO (dotdashed ) and NNLO (dotted)~\cite{GehrmannDeRidder:2007hr} and results for the PMC  scale-setting (solid line) are shown for $\sqrt{s}=35$, $44$ GeV respectively.
Error bands for the conventional predictions are obtained by varying $\mu_r\in[\sqrt{s}/2,2\sqrt{s}]$. Error bands for PMC predictions are given by the uncertainty $\Delta\alpha_s(M_Z)=\pm0.0009$~\cite{ParticleDataGroup:2024cfk}. The experimental data are taken from JADE~\cite{MovillaFernandez:1997fr,Biebel:1999zt}.}.
\label{eventshapeCPMC35}
\end{figure}

Moreover, we report results for the case of lower energies in Figs.(\ref{eventshapeTPMC35}) and (\ref{eventshapeCPMC35}). Here, we also show the comparison between the conventional method and PMC results at the center-of-mass energy of $\sqrt{s}=35$ GeV, and at $\sqrt{s}=44$ GeV. Experimental data for thrust ($T$) and C-parameter ($C$) at these energies have been measured by JADE~\cite{MovillaFernandez:1997fr,Biebel:1999zt} and DELPHI~\cite{Abdallah:2003xz}.
In Figs.(\ref{eventshapeTPMC35}) and (\ref{eventshapeCPMC35}), we notice that conventional results are also plagued by large renormalization scale uncertainties in the low-energy range.

As in the previous cases, both thrust ($T$) and C-parameter ($C$) results obtained with the conventional approach, show to underestimate data, while the use of the PMC scale-setting improves over a wide range of values of the selected event shape variable, the comparison with experimental data.
Comparing all results, it is noted that PMC predictions for the distributions above the $Z^0$ peak shown in Figs.(\ref{eventshapeTPMC}) and (\ref{eventshapeCPMC}), have a better agreement with the experimental data. This can be attributed to the impact of non-perturbative effects on event shape variables at low center-of-mass energies, that reduces while energy is increased, damping with a $1/E$-rate. Therefore, in order to obtain an improved description of the results, it is crucial to consider also the effects of non-perturbative terms on low-energy data.

\section{Summary}
\label{sec:4}

One way to test perturbative QCD and accurately determine the strong coupling constant $\alpha_s$ is to analyze the event shape observations in electron-positron annihilation. Careful study of the event shape observations is of great importance for understanding the strong interaction and extracting the coupling constant from experimental data. The conventional practice of guessing the scale often results in scale fuzziness, introducing significant uncertainty in theoretical predictions. Unlike conventional methods, the PMC method eliminates the uncertainties associated with renormalization scheme and scale ambiguities.
In this paper, we provide a detailed PMC analysis of the event shape variables in the $V$ scheme. We demonstrate that the PMC scales under the $V$-scheme exhibits a faster increase compared to the $\overline{\rm MS}$ scheme. Thus, a better convergence in the perturbative series can be obtained. Moreover, the PMC scales dynamically change with the value of the event shape observables, reflecting the virtuality of the quark and gluon subprocess. This dynamic change ensures that the PMC scale is consistent with the correct physical behavior of the scale, especially in the two-jet region. As a result, the event shape observables are in good agreement with experimental data, and for the case of the jet transition variable $Y_3$, we obtain a first improvement in the results to some extent compared with the $\overline{\rm MS}$ scheme. Furthermore, we also analyse the event shape distributions above and below the $Z^0$ peak. Event shape distributions above the $Z^0$ peak align well with experimental data, while PMC results below the $Z^0$ peak show significant enhancement compared with the conventional results. This is can be considered an effect of the non-perturbative terms that play an important role on the low-energy behavior. This analysis shows the importance of the application of a reliable renormalization scale setting method in perturbative QCD and we expect the application of the PMC methodology to be extended also to other fundamental processes, such as the event shape variables in electron-proton, proton-antiproton or proton-proton collisions.

\hspace{1cm}

{\bf Acknowledgements}: This work was supported in part by the Natural Science Foundation of China under Grant No.12265011, No.12175025 and No.12147102; by the Project of Guizhou Provincial Department under Grant No.YQK[2023]016, ZK[2023]141 and GZMUZK[2022]PT01; by the Hunan Provincial Natural Science Foundation with Grant No. 2024JJ3004, YueLuShan Center for Industrial Innovation (2024YCII0118).

\end{document}